\documentstyle[12pt]{article}

\def\bA{{\bar A}}
\def\i{{\frac1{2\pi}\int d^2z\,}} 
\def\R{{\cal R}e}
\def\Ri{{\frac{\R}{2\pi}\int d^2z\,}} 

\def\iv#1{{\frac{#1}{2\pi}\int d^2z\,}}
\def\d{\partial}
\def\bd{\bar\partial}
\def\bD{\bar D}
\def\a{\alpha}
\def\b{\beta}
\def\cD{{\cal D}}
\def\cA{{\cal A}}
\def\th{\theta}

\def\de{\delta}

\def\l{\lambda}
\def\s{\sigma}

\def\z{\zeta}
\def\bg{{\bar g}}

\def\sp{\; \; \; }
\def\t{\tilde}
\def\to{\rightarrow}
\def\implies{\Rightarrow}
\def\LR{\Leftrightarrow}
\def\dg{\inv g \d g}
\def\bdg{\inv g \bd g}
\def\sss{\scriptscriptstyle}
\def\spp{\sp\sp\sp\sp\sp }
\begin{document}

\newcommand{\inv}[1]{{#1}^{-1}} 

\renewcommand{\theequation}{\thesection.\arabic{equation}}
\newcommand{\beq}{\begin{equation}}
\newcommand{\eeq}[1]{\label{#1}\end{equation}}
\newcommand{\ber}{\begin{eqnarray}}
\newcommand{\eer}[1]{\label{#1}\end{eqnarray}}
\newcommand{\brr}{\begin{equation}\begin{array}}
\newcommand{\err}[1]{\end{array}\label{#1}\end{equation}}

\begin{center}
       July, 1997
                                \hfill   
                                \hfill   
                                \hfill  \\
                                \hfill  \\

\vskip .3in
{\large \bf A Change of Variables to the Dual \\ and \\
Factorization of Composite Anomalous Jacobians }
\vskip .4in

{\bf Eliyahu Greitzer}\footnote{e-mail address: eliyahu@astro.huji.ac.il}
\vskip .1in

{\em Racah Institute of Physics, The Hebrew University\\
  Jerusalem, 91904, Israel} \\

\vskip .30in

\vskip .1in
\end{center}
\vskip .2in
\begin{center} {\bf ABSTRACT } \end{center}
\begin{quotation}\noindent

Changes of variables giving the dual model
are constructed explicitly for $\s$-models without isotropy.  In particular,
the jacobian is calculated to give the known results.  The global aspects of the abelian case as well as some of those of the cases where the isometry group is simply connected are considered.\\
	Considering the anomalous case, we infer by a consistency argument that
the `multiplicative anomaly' should be replaceable by adequate rules for
factorization of composite jacobians.  These rules are then generalized in
a simple way for composite jacobians defined in spaces of different types.  Implimentation of these rules then gives specific formulas for the anomally for semisimple algebras and also for solvable ones.  
\end{quotation}
\vfill
\eject

\def\baselinestretch{1.2}
\baselineskip 16 pt

\section{Introduction}
\setcounter{equation}{0}
\setcounter{footnote}{0}

One of the striking features of string theory is target-space duality
\cite{GPR}.  This duality relates space-times of very different nature, that
correspond (locally) to the same CFT.  In particular, considering a
dualization with respect to a non-abelian isometry, the corresponding symmetry
admitted by the dual model is non-local, and exists no more as an
isometry.  From that, one may conclude the non-local nature of possible
transformations between the two models.  As suggested by Giveon, Rabinovici and
Veneziano \cite{GRV}, and proved by \'{A}lvarez, \'{A}lvarez-Gaum\'{e} and
Lozano \cite{AAL}, a $\s$-model admitting an isometry with vanishing isotropy
is related to its dual by a canonical transformation.\\

	In the following we derive an explicit change of variables that
produces the dual action for models without isotropy.  Dealing with
the jacobian, we resort to factorization rules for composite and inverse
operators' `determinants'.  The structure of the paper is as follows: In
section {\bf 2} we present and prove a general change of variables relating (locally) the
case without isotropy to its dual, classically; the corresponding jacobian is
produced in section {\bf 3}, relying on results from section {\bf 4} (which is
somewhat independent), where general rules for decomposition of determinants of
composite tensors are given.  The rules are inferred by the requirement that
functional changes of variables be consistent.  The global aspects of the case where the isometry group is abelian and the space is (possiblly) curved can be found in section {\bf 5}.  Still in that section -  some global aspects of the non-abelian simply-connected case are realized.  Section {\bf 6} is dedicated to reviewing some of the above results and their significance. 
\section{The Change of Variables}
\setcounter{equation}{0}
	The main model regarded in this paper is the general case without
isotropy studied in \cite{GR},\cite{EGRSV} and presented here briefly.
Consider a target space with coordinates $g$ that transform as $g\rightarrow
ug$ where $u,g\in G$, ($G$ is some Lie group), and further inert coordinates
$x^i$.  The general action can be written in the form
\ber
S[\dg,\bdg,x]\; = \;
\i \Big( (\dg)^a E_{ab}(x)(\bdg)^b &+ \nonumber \\
(\dg)^a F^R_{aj}(x)\bd x^j + \d x^i F^L_{ib}(x)(\bdg)^b
+ \d x^i F_{ij}(x) \bd x^j &- \nonumber \\ \Phi (x)\d\bd\s\Big)\ ,
\eer{original}
where $\s$ is the conformal factor and
\beq
(\inv g\d g)^a \equiv tr (\t{T}^a \inv g\d g) \ \ \LR \ \ \inv
g\d g = (\inv g\d g)^a T_a\ ,\ \ etc.
\eeq{gdga}
The generators $T_a$, $a=1,...,dim(G)$, obey
\beq
[T_a,T_b]=f^c_{ab}T_c \sp\sp ,
\eeq{TTfT}
and the `dual generators' $\t{T}^a$ are defined by the condition
\beq
tr(T_a \t{T}^b)= \delta_{a}^b \sp\sp .
\eeq{trTT}

To construct the dual model (for review see \cite{GPR} and references therein),
one gauges (minimally) the isometry group with gauge fields $A, \bA$ (in
complex worldsheet coordinates).  These fields are then constrained to be flat
by the addition of the term $\l_{c}F^{c}(A,\bA)$ to the lagrangian, upon
integrating out of the lagrange multipliers $\l_{c}$.  Gauge fixing $g=1$ then
gives the action 
\beq
S[A,\bA,x] + \i \l_{c} (\d \bA^{c} - \bd A^{c} + Af^{c}\bA ) \sp ,
\eeq{fixed}
where $S[A,\bA,x]$ is (\ref{original}) with $\dg$ and $\bdg$ replaced by the
{\em independent} fields $A$ and $\bA$, respectively.
({\bf Note:} from now on matrix and vector indices will sometimes be
supressed).  Finally, (gaussian) integrations over the gauge fields yield the
form of the {\em dual} model (in the non-anomalous case)
\ber
S_{dual}[\l ,x] &= & \i \Big( (\d \l _a-\d x^iF^{L}_{ia})
N^{ab}(\bd\l_b+F^{R}_{bj}\bd x^j)  \nonumber \\ & + &
\d x^i F_{ij}\bd x^j - (\Phi -ln \det N)\d \bd \s \Big) \ ,
\eer{dual}
where
\beq
N^{ab}\: \equiv \: [ (E+ f^{c} \l_c)^{-1}]^{ab}\sp .
\eeq{N}In the anomalous case, the only correction to (\ref{N}) is an extra
non-local term proportional to $tr T_{a}$
(\cite{GR},\cite{ALVY},\cite{EGRSV}).\\

	In this note we present another way to derive the dual model.  Namely\,
we perform a change of variables in the functional integral from $\{ g,x\}$ to
$\{ \l ,x\} $ 
\brr{ccccc}
a)&(\dg )E(x)\ &=& \d \l - \d x  F^{L}(x)-(\dg )  f^{c} \l_c &\LR \\
b)&  \dg  &=& (\d \l - \d x F^{L}(x))(E(x)+f^{c}\l _{c})^{-1}& 
\err{Generaltrans}
~\footnote{If
the structure constants are totally anti-symmetric (i.e. for compact
semi-simple group) one also has $\d (\inv g \l g) = ((\inv g\d g)^a E_{ab}  +\d
x F^L_{b})\inv g T^b g$, from which $\l$ is derived explicitly.}  
(see section 5 for the change of variables of opposite chirality). \\
The proof for that runs as follows:\\

	Substituting \ref{Generaltrans}$a$ in the first term of
(\ref{original}), using
\beq
F[\dg,\bdg] = 0 \sp\sp ,
\eeq{flat}
and finally substituting \ref{Generaltrans}$b$ one gets the identity
\ber
(\dg)^a E_{ab}(x)(\bdg)^b +(\dg )^a F^R_{aj}(x)\bd x^j + \d x^i F^L_{ib}(x)(\bdg)^b \d x^i & &= \nonumber \\ 
(\d \l _a-\d x^iF^{L}_{ia}) [(E+ f^{c} \l_c)^{-1}]^{ab}(\bd\l_b+F^{R}_{bj}\bd
x^j)\; + \; \bd(\l\inv g \d g)-\d(\l\inv g \bd g).
\eer{identity}
Equation (\ref{identity}) relates an action in group variables to an action in
the algebra variables, up to a total derivative term which is discussed in
section 5.  This completes the proof in the level of the lagrangian.\\

	Next, we turn to study the jacobian for the transformation
(\ref{Generaltrans}).
\section{The Jacobian}
\setcounter{equation}{0}
Denoting both sides of (\ref{Generaltrans})$b$ as $\cA$, their variations with
respect to $g$ and $\l$ are respectively
\brr{lcccc}
LHS  \sp & & D^{c}(\inv g \de g)& =& \d (\inv g \de g)^{c} + \cA ^{a}
f_{ab}^{c} (\inv g \de g)^{b} \\
RHS \sp & & \left( -\t{N} \t{D}(\de \l)\right) ^{c} &= & N^{bc} (\d \de \l_{b} -\cA
^{a}f_{ab}^{d}\de \l _{d})\sp ,
\err{variation}
where  $\:\t{}\:$  denotes the functional transpose~\footnote{Notice the
similarity between the characterization of the symmetry $ D(\inv g \de g)=0 $ of
the original model - and the corresponding symmetry of its dual: $\t{D}(\de
\l)=0$.  See section 6 for a possible significance of such similarity.  The
(global) {\em gauge invariant measure} for $\l$ is given
in ($\ref{variation}_{RHS}$).  Its dependence on the background fields is due to
the non-local nature of the symmetry as we transform $g\to\l$.  Another thing 
worth mentioning at this point is that (\ref{Generaltrans}) also relates {\em
two different field equations} - that of the original model, which looks like:
$\d \bar J (g,x) +\bd J(g,x) =0$ and that of the dual - $F(A(\l ,x ),\bA (\l
,x))=0$ (see \cite{GR} for the exact forms).}.  The required jacobian is thus
\brr{ccccc}
J &=&\frac{\cD g}{\cD \l }&=& | D^{\sss -1}\t{N}\t{D}| \sp ,
\err{Jacobian}
where $|x|$ stands for $\det(x)$.  This can be proved to be the ratio
\ber
| N| \frac{| D\bD| }{|D ||\bD|}\Big| _{\bg =g }&=&| N|\exp \{ -\iv{tr}(\dg
\bd\s+\bdg\d\s )\}
\eer{NL}
calculated in \cite{EGRSV} to the first order in the conformal factor $\s$.
These determinants correspond to the changes $F(A(g),\bA (\bg ))\to g,g\to
A,\bg\to\bA$ (parametrized as in (\ref{A's})).  A general practice for treating
`determinants' of inverse and transpose operators is one of the offshoots of
the next section, where the equivalence of (\ref{Jacobian}) and (\ref{NL})
follows naturally. \\

	One might want to derive the general form of the dual model found in
\cite{EGRSV}.  To that end, notice that
\brr{ccrclc}
0 &=& tr \int\s F(\dg ,\bdg )&=&tr\int \s\d (\bdg ) -tr \int\s\bd (\dg
)&\implies \\
& & tr\int \bdg \d\s &=&tr\int \dg \bd\s &
\err{ganomaly}
so that by (\ref{Generaltrans}), (\ref{NL}) becomes
\ber
J &=& | N| \exp \{ -\iv{2}\bd\s N(\d \l - \d x F^{L})\} \sp ;
\eer{noname}
then, by substituting the equations of motion for $\l$ \cite{GR}, one obtains
the same form for the terms linear in $\s$ as in \cite{EGRSV}~\footnote{A quick
way to obtain the very result found in \cite{EGRSV} is by using the $\bf \s$
{\em dependent} transformation $\dg= (\d \l - \d x F^{L}+\d\s tr\: T)N$.}.
\section{Factorization of Jacobians}
\setcounter{equation}{0}
{\em The Motivation}\\  Consider some action $S$ that depends on the group
variables $g,\bg$ through the (gauge) fields
\beq
A =\dg \sp , \sp\sp \bA= \bar g^{-1} \bd \bar g \sp\sp ,
\eeq{A's}
which in turn appear in $S$ in combinations $F(A,\bA )=D \bA -\bd A$.  When
changing variable $\bg\to \bA$ followed by the change $\bA\to F$, the {\em
total} jacobians multiplying the partition function's integrand are
\ber
J& = &(| D | | \bD | )^{-1} \sp .
\eer{ratio}
On the other hand, changing $F \to \bg$ directly, one collects the jacobian
$|D \bD | ^{-1}$.
Now, for path integration to be consistent, the results of these two courses of
changing variables from $\bg$ to $F$ should be no different in the end.  But as
we saw in (\ref{NL}), their ratio is non-trivial, at least for groups with
traceful structure constants; this is {\em the mixed anomaly} \cite{ALVY}, here
in the form of a `multiplicative anomaly' \cite{EGRSV} which seems to violate
the functional chain rule.\\
{\em Factorization}\\
To resolve this puzzle~\footnote{The author wishes to thank S. Elitzur for
suggesting the direction which led to the formulation and also for the proof (\ref{nonmixed}).} the ghost actions (the
variation of which with respect to the conformal factor should give the value
of the anomaly) defining the functional determinants are invoked~\footnote{the
$\z$-function procedure is not defined for jacobians relating two spaces of
different types.  When both procedures are defined, one might want to prove
that they are different by local counterterms at most.}. Let us write the ghost
actions in interest:  we have $g,\; \bg$ and $F$ that are worldsheet scalars
where $A,\bA$ are components of a worldsheet vector; the fermionic ghosts of
types $s$ (for scalar) and $v$ (vector) are thus introduced.  Changing $A \to
g$, the corresponding jacobian may be written as~\footnote{Such definitions are
not Lorentz-invariant and are ill-defined in general \cite{ALVY}.  This,
however, should not interfere with our argument which is compelled by the chain
rule.} 
\brr{ccccc}
{\cD A}/{\cD g}&=&| D| _{vs}&\equiv & \int \cD v \cD s \exp {\int vDs}
\err{D}
where the integration in the exponent is over the worldsheet, partial
derivatives change to worldsheet covariant ones, and indices of all types are
supressed.  By partial integration we have
\brr{ccccccc}
\int vDs&=&\int s\t{D}v &\implies & | D| _{vs}&= &  | \t{D} | _{sv} \sp .
\err{a}
Further, bearing in mind the chain rule, one can factorize and re-merge
`determinants' of vector operators` products and derive identities such as:
\brr{cccccc}
| O_{1}O_{2}| _{ss} &= &| O_{1}| _{sv}  | O_{2}| _{vs} &= &| \t{O_{1}}|
_{vs}| \t{O_{2}}| _{sv} &= \\ | \t{O_{1}}\t{O_{2}}| _{vv}  &=& |
(O_{2}O_{1}\t{)}| _{vv}&=& | O_{2}O_{1}| _{vv}& \sp .
\err{ident1}
By considering a change of variables and its inverse change, we also have  
\brr{ccccc}
| O_{1}^{\sss -1}O_{2}| _{ss} &= & | O_{1}^{\sss -1}| _{sv}  | O_{2}| _{vs} &=&
| O_{1}| _{vs}^{-1}  | O_{2}| _{vs} \sp.
\err{ident2}

These rules for chaining jacobians are easily generalized to jacobians
of tensors relating two objects, possibly of different ranks (with respect to
worldsheet diffeomorphisms).  The basic rules for that are
\brr{cccc}
a)\sp &| A| _{r_{1}r_{2}}&=&| \t{A} | _{r_{2}r_{1}} \\
b)\sp &| A| _{r_{1}r_{2}} | B| _{r_{2}r_{3}}&=&| AB| _{r_{1}r_{3}}\\
c)\sp &| A^{\sss -1}| _{r_{1}r_{2}}&=&| A| ^{-1} _{r_{2}r_{1}}
\err{rules}
with obvious notations.   These rules should be correct {\em whenever they
correspond to legitimate changes of variables.} \\
Applying these rules to the anomaly (\ref{NL}), it may take the
following equivalent forms
\ber
\frac{|D \bD | _{ss}}{| {D}|  _{vs}| \bD|  _{vs} } = \frac{|{D}|_{sv}|\bD|
_{vs} }{| {D}|  _{vs}| \bD|  _{vs}}= \frac{|D | 
_{sv}}{ | D | _{vs}} =\frac{|D | _{sv}}{ | \t{D} | _{sv}} \nonumber \\ =  |
\t{D}^{\sss -1}| _{vs} | D | _{sv}= | \t{D}^{\sss -1} D | _{vv}= |
D\t{D}^{\sss -1}| _{ss}\sp .
\eer{forms}
This proves the equality of (\ref{Jacobian}) and (\ref{NL}) in particular.  As
to the example in the beginning of the section, the corresponding ratio is
\ber
\frac{|D\bD |_{ss}}{|D|_{sv}|\bD |_{vs}}&=&1
\eer{b}
according to (\ref{forms}) or (\ref{rules}).\\

	With (\ref{forms}) in mind, let us consider the anomaly in two classes of cases.  By definition, we have 
\ber
\int vDs&=&\int v_{b}\d s^{b} + v_{c}f^c_{ab}A^a s^b . 
\eer{D}
{\em If $G$ is semisimple,} then $f_{abc}$ is totally antisymmetric.  Reshuffeling indices and integrating by parts, (\ref{D}) may be written as
\ber
\int -s_{b}\d v^{b} - s_{b}f^b_{ac}A^a v^c=-\int sDv &\implies&|D|_{vs} =|D|_{sv} 
\eer{}
$\implies$ no anomaly, by virtue of (\ref{forms}).\\

	{\em If $G$ is solvable}, there exists a triangular basis for the algebra s.t.
\brr{rll}
\int vDs=\\ \nonumber
\int & v_1\d s^1 +A^\mu f^1_{\mu 1} v_1 s^1 \\ \nonumber
\sp +&v_2\d s^2 +A^\mu f^1_{\mu 2} v_1 s^2 +A^\mu f^2_{\mu 2} v_2 s^2 \\ \nonumber
&\sp.\spp . \spp\sp\sp.\\ \nonumber
&\sp.\spp . \spp\spp. \\ \nonumber
&\sp.\spp . \spp\spp\sp\sp\sp. \\ \nonumber
\sp +& v_N\d s^N +A^\mu f^1_{\mu N} v_1 s^N+\sp.\sp.\sp . \sp+A^\mu f^N_{\mu N} v_N s^N\: . 
\err{D2}

Integrating over $s_1$ and then over $v_1$, produces the functional determinant
\ber
|-\d+A^\mu f^1_{\mu 1}|_{sv}\sp ,
\eer{2.5}
while setting $v_1$ to zero, by which all of the terms in the second column vanish.  Repeating this procedure for $(s^2 , v_2),...,(s^N , v_N)$, we finally get the formula
\ber
| D|_{vs} &=& \prod_{k=1}^N |-\d+A^\mu f^k_{\mu k}|_{sv}=\prod_{k=1}^N |\d\;{\mathbf -}\; A^\mu f^k_{\mu k}|_{sv}\sp .
\eer{DE}
Switching $s$ and $v$, we get
\ber
&| D|_{sv}\sp =\sp \prod_{k=1}^N |-\d+A^\mu f^k_{\mu k}|_{vs}=\prod_{k=1}^N |\d-A^\mu f^k_{\mu k}|_{vs}\sp=&\nonumber\\ &\prod_{k=1}^N |\d\;{\mathbf +} \; A^\mu f^k_{\mu k}|_{sv}\sp ,&
\eer{DD}
so the anomaly , which is the ratio of (\ref{DE}) and (\ref{DD}) can be written as a product of chiral anomalies.  By Adler and Bardeen \cite{ADBAR}, we conclude that if the anomaly vanishes to first order, it cancels altogether.  The condition for that is \cite{EGRSV} 
\ber
\sum_k f^k_{\mu k}&=&0\sp .
\eer{D5} 
The methods above can also be used for the general case, i.e. a semi-direct product of a semisimple group and a solvable one (e.g. the Lorentz group).  However, the general classification of such groups is still a mystery and so is the general rule for factorization of the corresponding covariant derivatives.


\section{Global Aspects}
\setcounter{equation}{0}
{\bf Notations and mathematical tools on compact Riemann surfaces}\\
{\bf 1)}  The $z$-component of a one-form $\omega =(\omega_z,\omega_{\bar z})$ can be completed to a full {\em closed} singled-valued one-form of which $\omega_z$ is its $z$-componet, as follows\ber
\t \omega =(\omega_z,\t \omega_{\bar z})= (\omega_z,\bd\int_{z_0} dz\, \omega_z) = dC\sp\sp (z_0 =const)\sp ,
\eer{complete}
where $C$ is a multivalued function on the surface.  One can verify the above on the torus and therefore on every handle of the surface.\\  
{\bf 2)}  For a closed one-form $Y$, let $Y_{0}$ stand for its zero mode(s) where $Y_{e}$ denotes its exact part, that is, $Y_0\equiv\sum_j \a_j\oint_{\aleph_j} Y$ and $Y_e \equiv Y-Y_0$, where $\{\a_j\}, j=1,...2{\bf g}$ is the (unique) basis to the space of harmonic differentials on a surface of genus $\bf g$, satisfying $\oint_{\aleph_j} \a_k= \delta_{jk}$ and $\{\aleph_j\}$ is a basis to the first homology group of the surface.\\
{\bf 3)}  For two closed one-forms $\a $ and $\b $ one has 
\ber
\int \a\wedge \b &=& \sum _{i=1}^{n} \oint_{a_i}\a\oint_{b_i}\b
-\oint_{b_i}\a\oint_{a_i}\b \sp ,
\eer{riemann}
where $\{a_{i},b_{i}|i=1,...,{\bf g}\}$ denote the non-trivial cycles on the
$\bf g$-genus surface \cite{Farkas}.\\ 
{\bf 4)}  If $\th (z,\bar z )$ and $\l ( z, \bar z )$ are two multivalued functions on a surface one has
\ber
\int d^2z\,(\d \th )_0 (\bd \l )_e =\int d^2z\, \bd ((\d \th )_0\l_s ) -\int d^2z\, \bd(\d \th )_0\l_s =0 -0 =0\sp ,
\eer{nonmixed} 
where the subscript $s$ stands for the single valued part, that is, 
\ber
\l_s (z, \bar z) = \int_{z_0}^z \,(d\l)_e\sp,
\eer{}
and thus $ (d\l)_e = d\l_s$.  We deduce\ the decoupling of terms, that are a product of zero modes and  exact modes - from the action.\\
{\bf 5)}  The equations of motion for the zero modes of a one-form ${\mathbf A}^a=(A,\bA)^a$ in actions of the form
\ber
{\mathbf S}&=&\i  (A M\bA + A N +K\bA)  
\eer{}
where the matrices $M,N$ and $K$ don't depend on $A$, {\em may be} written as
\ber
(M\bA +N)_0&=&( AM +K)_0\sp =\sp 0\sp .
\eer{zmeof}
The zero modes are defined by {\bf 2)} after completing $N$, $K$, $A$ and $\bA$ ({\bf 1)}) and using {\bf 4)} for the decoupling of the exact and the zero modes. \\
{\tt An abelian case}\\
{\em The original model}:  Let the original action be
\ber 
S_{original}&=&\i\d\th R^{2}\bd\th\sp ,
\eer{abeloriginal}
with $R=R(x(z,\bar z))$, $\th =  \th +2\pi l\sp ,l=const$.  In order to dualize (\ref{abeloriginal}) we write an equivalent action
\ber
S&=& \Ri\d\th R^{2}\bd\th\sp +\sp(\d\th)_0(\bd\l)_{0}-(\d\l)_{0}(\bd\th)_0\sp ,
\eer{start}
where now $\th$'s holonomies are any $2{\bf g}$ real numbers and $\l$ (who's single valued part is yet unspecified) satisfies
\ber
\oint_{a_i}(d\l)_0 =\frac{4\pi m_{i}}{l} \;,\sp\oint_{b_i}(d\l)_0 =\frac{4\pi n_{i}}{l}\sp .  
\eer{periodicities}      
To recover $\th$'s original periodicity, we write the total derivative term in (\ref{start}) as $\frac{i}{4\pi}
\int (d\th)_0\wedge (d\l) _0$, which by (\ref{riemann}) equals $\frac{i}{l}\sum_{j}(n_{j} \oint _{a_{j}}d\th -m_{j} \oint _{b_{j}}d\th)$. {\em Notice that the zero subscripts in} (\ref{periodicities}) {\em are omittable}.  Summing over all possible integer $n_{j}$ and $m_{j}$, constrains $\th$'s periodicity to be $2\pi l$.\\
{\em The dual model}:  Imposing the equations of motion of $(\d \th)_0$ and $(\bd \th)_0$ on (\ref{start}) imposes
\ber
(\d\l)_0 - (\d\th R^2)_0&=&0\sp .
\eer{abelglobal}
Therefore, by virtue of {\bf(1)} we define
\ber
(\d\l)_e&=&(R^2\d\th)_e \sp ,
\eer{eabelcov}
and use (5.10) to write
\ber
\d\l&=&R^2\d\th
\eer{abelcov}
and substitute it in the first term of (\ref{start}), which after omitting the zero subscripts form its next terms takes the form   
\ber
&\Ri\d\l\bd\th +\d\th\bd\l-\d\l\bd\th =\Ri\d\th\bd\l&\stackrel{\d \th=\d\l R^{-2}}{=}
\eer{} 
\ber
&S_{dual}\sp=\sp\Ri\d\l R^{-2}\bd\l  \sp , &
\eer{abeldual}
where $\l=\l+\frac{4\pi}{l}$.\\  This can readily be generalized to (\ref{original}) with $G=U(1)^{dimG}$.  To omit the $\R$ in (\ref{abeldual}), one may change variables by replacing an integrand of the same form, with a real $\l$.  The realness issue carries over to the non-abelian case and so should its solution.\\
{\tt A non-abelian case}:  Let the original action be
\brr{ccc}
a)&\sp S = \i(\dg)E(\bdg) =\\ \nonumber 
b)&\sp \i((\dg)E)E^{-1}(E(\bdg)) = \\ \nonumber
c)&\sp tr\Ri(g(\dg) Eg^{-1})_{0}(\bd g g^{-1})_{0}+(g(\dg) Eg^{-1})_{e}(\bd g g^{-1})_{e}\sp ,  
\err{Nstart}
with $G$ simply connected and $(\dg)E \stackrel{def}{=}((\dg)E)_b T^b\stackrel{def}{=}(\dg)^{a}E_{ab}T^b\;$, etc.  To derive  (\ref{Nstart}c) from (\ref{Nstart}a), we have used the trace cyclicity along with the conclusion from {\bf 4)}.\\
{\em The dual model}\,\, Since all mappings from a Riemann surface to a simply connected group are homeotopic to a point (on the group's manifold), the configurations space of $g$ is continuous and connected, therefore we can invoke the equations of motion of 
$(Eg^{-1}dg)_0$ together with {\bf5)}, to constrain $g$'s configurations s.t.
\ber
( g^{-1} dg)_0 &=& 0\sp ;
\eer{zm}
{\em the extent of validity of this step is yet to be examined (elsewhere)}.  Substituting $g$ with $g^{-1}$ we can characterize this subspace also by 
\ber
(gdg^{-1})_0 = - (dg g^{-1})_0 & = & 0\sp ,
\eer{zm1} 
so that the first term in (\ref{Nstart}c) decouples from the action.
Using {\bf 1)} we define the function $\chi$ by
\ber
\d\chi&=& g(\dg) Eg^{-1}\sp .
\eer{ONcov}
Notice that $\chi$'s possible multivalued part decouples from the action and may therefore be taken to be zero.  We substitute (\ref{ONcov}) in what's left of 
(\ref{Nstart}) to give
\ber
tr\Ri\d\chi \bd g g^{-1}\sp.
\eer{c}
Integrating it by parts and then using the identity $\d(\bd g g^{-1})=g\bd (g^{-1}\d g)g^{-1}$ yield
\ber
-tr\Ri\chi g\bd (g^{-1}\d g)g^{-1}\sp .
\eer{b}
Then, using the trace cyclicity and integrating by parts again we obtain
\ber
tr\Ri\dg\bd\l\sp ,
\eer{a}  
where $\l=g^{-1}\chi g$ is single valued.  Finally, we observe that (\ref{ONcov}) may equivalently be written as 
\ber
\inv g \d (g\l\inv g)g = (\dg)E \LR  \dg\l+\d\l-\l\dg = \dg E \LR\nonumber \\
\d\l=\dg(E+f^c\l_c)\LR\dg=\d\l(E+f^c\l_c)^{-1}\; ,
\eer{e}
from which we obtain
\ber
S_{dual}&=&\i\d\l(E+f^c\l_c)^{-1}\bd\l\sp ,
\eer{Ndual}
where {\em the $\l$'s are single valued.}  The total derivative term from the two integrations by parts 
\ber
tr\frac{i}{2}\Ri d(\chi dg\inv g)
\eer{d}
vanishes because that $\chi dg\inv g$ is single valued.  The generalization to (\ref{original}) is again, straightforward.

\section{Concluding Remarks}
The constructions presented in this paper imply (once again) that in the case
without isotropy, the conformal invariance of an action and that of its dual
are equivalent {\em to all orders of} $\a '$ (assuming a correct computation of
the jacobian, especially if that is non-local).\\

	The author's hope is that the change of variables presented in this
work will shed some light on the global issues in dual models of isometry groups the mappings to which from Riemann surfaces fall into more interesting homology structure than the one presented here i.e. the trivial one that corresponds to simply-connected groups. \\

	The (classical) equality in the case without isotropy
(\ref{identity}) between the original and the dual action
might seem a surprise, since the former admits a symmetry that seemingly
is absent in the latter \cite{OQ}. Further, a check of the change of variables
(\ref{Generaltrans}) verifies that transforming $g \to ug$ induces no change in
the dual coordinate $\l$.  Where has the original symmetry gone? It has gone
non-local (cf.\cite{GR}) just to avoid detection by the Killing
equation~\footnote{With (\ref{Generaltrans}) in mind, one suspects the
existence of some non-local generalization of the Killing equation that is
capable of detecting such hidden symmetries, thus providing means by which
dualization may be reversed.}. Actually, the very condition for the
`smoothness' of the change of variables, that is, for the jacobian to be local,
is that the {\em symmetry factors} (see footnote after eq. (\ref{variation}))
$|D |$ and $| \t{D} |$ before the volume elements $\cD g$ and $\cD \l$
respectively should cancel out.  It is precisely when this correspondence
between the symmetries breaks, that the anomaly occurs.   \\

	As shown in section {\bf 4}, the factorization approach to reorganize composite jacobians may be of help in tracing and isolating the very generators of the anomaly, as well as in simplifying its ghost action. One might want to generalize that to all Lie algebras and prove the chiral nature of the mixed anomaly in general.\\

\vskip .2in \noindent
{\bf Acknowledgements} \vskip .2in \noindent
I would like to thank my supervisor A. Giveon for his guidance and for
essential help in preparing this paper.  I acknowledge S. Elitzur for very
useful discussions, and thank A. Babichenko, S. Forste, O. Pelc and
G. Sengupta for their help.  Special thanks to G. Spivak for her help and
support.  This work is supported in part by BSF -
American-Israel-Binational-Science-Foundation and by the BRF -
Basic-Research-Foundation.

\end{document}